\documentclass{emulateapj}
\usepackage{rotating,psfig,natbib}

\def\Rjup{$\mbox{R}_{Jup}$}
\def\Rlam{$\mbox{R}_{\lambda}$}
\def\h2o{H$_2$O}

\def\hd{\mbox{HD209458b}}

\def\simgr{\,\hbox{\hbox{$ > $}\kern -0.8em \lower 1.0ex\hbox{$\sim$}}\,}
\def\simle{\,\hbox{\hbox{$ < $}\kern -0.8em \lower 1.0ex\hbox{$\sim$}}\,}

\begin{document}
\bibliographystyle{apj}

\title{Identification of Absorption Features in an Extrasolar Planet Atmosphere }
\author{T. Barman}
\affil{Lowell Observatory, 1400 W. Mars Hill Rd., Flagstaff, AZ 86001, {\tt barman@lowell.edu}}

\begin{abstract}
Water absorption is identified in the atmosphere of HD209458b by comparing
models for the planet's transmitted spectrum to recent, multi-wavelength,
eclipse-depth measurements (from 0.3 to 1 $\mu$m) published by
\cite{Knutson07}.  A cloud-free model which includes solar abundances, rainout
of condensates, and photoionization of sodium and potassium is in good
agreement with the entire set of eclipse-depth measurements from the
ultraviolet to near-infrared.  Constraints are placed on condensate removal by
gravitational settling, the bulk metallicity, and the redistribution of
absorbed stellar flux.  Comparisons are also made to the \cite{Charb02} sodium
measurements.   
\end{abstract}

\keywords{planetary atmospheres - extrasolar planets}

\section{Introduction}
The discovery of transiting extrasolar
planets has opened the door to direct detections and characterization of their
atmospheres.  Observations using the STIS instrument on HST provided the first
glimpse of what the photospheric composition is like for a nearby EGP.
\cite{Charb02} measured the relative change in eclipse depth for HD209458b
across a sodium doublet (5893\AA) resulting in the first detection of atomic
absorption in an EGP atmosphere.  Following the Na detection, \cite{Madjar03}
discovered an extended hydrogen-rich atmosphere surrounding HD209458b using a
similar technique as Charbonneau et al., but in the UV.  At Lyman-$\alpha$
wavelengths, HD209458b is $\sim 3$ times larger than in the optical.  These
detections were made using a technique called transit spectroscopy as the
planet passes in front of its star.  Transit spectroscopy uses the fact that
the wavelength-dependent opacities in the planet's atmosphere obscure stellar
light at different planet radii leading to a wavelength-dependent depth of the
light-curve during primary eclipse.  Consequently, searching for relative
changes in eclipse depth as a function of wavelength directly probes the
absorption properties of the planet's atmosphere with the potential to reveal
the presence (or absence) of specific chemical species.

In this paper, recent measurements of HD209458b's radius are combined in a
multi-wavelength comparison to model atmosphere predictions.  Atmospheric
molecular and atomic absorption are identified and constraints are placed on the
basic atmospheric properties.

\begin{figure*}[!thb]
\plotone{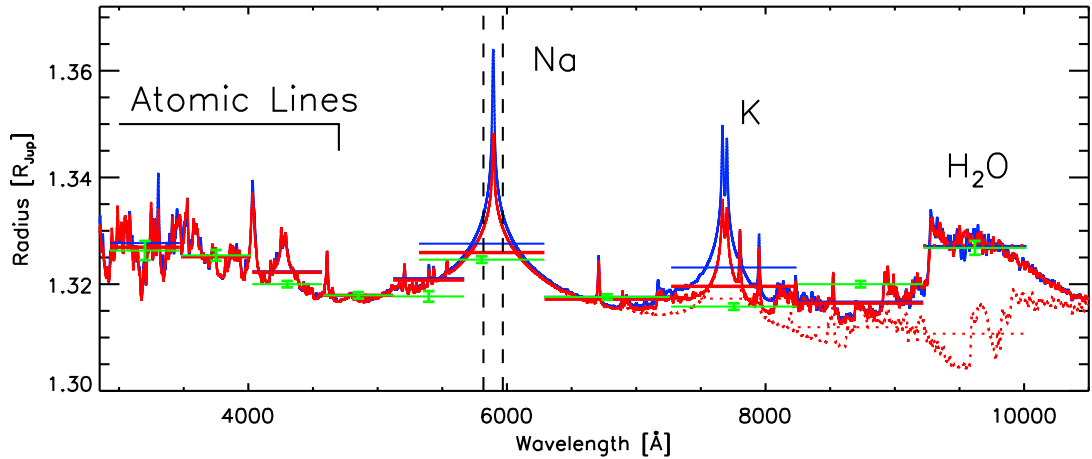}
\caption{
Monochromatic transit radii over the STIS spectral range for the baseline
rainout model with (red) and without (blue) photoionization of Na and K.  The
solid and dotted red lines are the same, except H$_2$O line opacity is excluded
in the latter. Horizontal bars correspond to mean radii across bins with
$\lambda$-ranges indicated by the width of each bar.  STIS measurements by
Knutson et al.  (2007) are shown in green with 1 $\sigma$ error bars. Vertical
dashed lines mark the narrow $\lambda$-range used by \cite{Charb02}.
\label{rlam1}}
\end{figure*}

\section{The Limb Model}

Transit spectroscopy probes the {\em limb} of a planet which is the transition
region between the day and night sides.  One would, therefore, expect that, in
the presence of a horizontal temperature gradient between the heated and
non-heated hemispheres, the temperatures across the limb would be cooler than
the average dayside temperatures \cite[]{Barman05,Iro05}.  Recent Spitzer
observations of both transiting and non-transiting hot-Jupiters showing large
flux variations with phase have provided strong evidence supporting such a
day-to-night temperature gradient \cite[]{Charbonneau05, Deming05, Deming06,
Harrington06}.

Describing the limb ultimately requires a multi-dimensional model atmosphere
solution; however, as is common practice, a simpler one-dimensional model is
used here to represent the average properties of the limb, in both a
longitudinal and latitudinal sense.  To explore a variety of limb temperature
structures, the incident stellar flux has been scaled by a parameter $\alpha$.
A model with $\alpha = 0.25$ represents an average description of the entire
planet in the presence of very efficient horizontal energy
transport \cite[]{Barman05}.  Increasing $\alpha$ to 0.5 increases the heating of
the model atmosphere making it more appropriate for just the {\em dayside}
which, in this work, will represent an upper limit to the plausible mean
temperatures at the limb.
 
The basic chemistry across the limb is modeled assuming chemical equilibrium,
determined by minimizing the free energy while including grain formation.  To
understand the effects of gravitational settling of grains on the chemistry and
the transmission spectrum, the removal of grains from the atmosphere is
included via two simple approximations that represent opposite extremes.    The
first is the ``cond'' approximation used by \cite{Barman01} and \cite{Allard01}
which simply ignores the grain opacity without altering the chemistry or
abundances.  The second is the ``rainout'' approximation which iteratively
reduces, at each layer, the elemental abundances (by the appropriate
stoichiometric ratios) involved in grain formation and recomputes the chemical
equilibrium with each new set of stratified elemental abundances.  This
approach is similar to other rainout models \cite[]{Fegley96,Burrows1999},
except that the depletion of elements is continued until grains are no longer
present.  

The transmission of stellar fluxes through the limb of HD209458b is determined
by solving the spherically symmetric radiative transfer equation while fully
accounting for scattering and absorption of both intrinsic and extrinsic
radiation \cite[]{Hauschildt92, Barman01, Barman02}.  Spherical, instead of the
more traditional plane-parallel geometry, naturally accounts for the curvature
of the atmospheric layers and changes in chemistry along the slant paths
through the upper atmosphere.  The planet radius at a given wavelength (\Rlam)
is obtained by determining the radial depth at which the transmitted flux is
equal to $e^{-1}$ times the incident starlight along that same path.  

\section{Results}
A cloud-free atmosphere with rainout, $\alpha=0.25$, and solar abundances is
adopted as the baseline model.  This model, along with others, is compared to
the relative \Rlam\ measurements of \cite{Knutson07} which have a reported
precision high enough to constrain many of the basic atmospheric properties.
Since a comparison is being made to {\em relative} \Rlam\ values, the model
results were uniformly scaled to match the observations in the 4580 to 5120\AA\
wavelength bin; this scaling was always less than 0.005\Rjup.  Overall, the
baseline model (red solid line in Fig. \ref{rlam1}) reproduces the observed
rise in \Rlam\ toward shorter wavelengths, the increase across the Na doublet,
and the increase at the far red wavelengths. The baseline model comparison to
the data has a $\chi^2$ that is 3 times {\em smaller} than a constant \Rlam.

\subsection{Water Absorption} 
Water is predicted to be one of the most abundant species in an EGP atmosphere
and, given its broad absorption features in the infrared, plays a crucial role
in regulating the temperature-pressure (T-P) profile.  The first major \h2o\
absorption band appears between 0.8 and 1 $\mu$m, a region covered by the last
two wavelength bins of \cite{Knutson07}.  As illustrated in Fig. \ref{rlam1},
there is excellent agreement between the baseline model and the observations in
this part of the spectrum especially across the longest wavelength bin that
sits on top of the \h2o\ band.  Qualitatively similar water features
are seen in the models of \cite{Brown01} and \cite{Hubbard01}; however these
models fall many $\sigma$ below the observations.  The baseline model
also predicts mean \Rlam-peaks equal to 1.330, 1.343, and 1.341
\Rjup\ for the next three water bands (at $\lambda \sim 1.15$, 1.4, and
1.9 $\mu$m).

A model that excludes \h2o\ line opacity is also shown in Fig. \ref{rlam1} and
is greater than 10 $\sigma$ below both the observations and the baseline model
prediction.  Removing \h2o\ opacity also produces a significant drop in \Rlam\
near 0.9 $\mu$m that further increases the discrepancy between the model and
overall red/near-IR observations.  No other opacity source could be responsible
for the observed rise in \Rlam\ across this part of the spectrum.    

\subsection{Photoionization}
After the reported sodium detection in the atmosphere of \hd\ \cite[]{Charb02},
there were several attempts to explain why the strength of this feature was
much lower than expected based on earlier models (e.g., by Seager \& Sasselov,
2000\nocite{Seager2000a}).  \cite{Barman02} explored departures from local
thermodynamic equilibrium (LTE) which are capable of producing an inversion in
the cores of the Na doublet line profiles.  However, the earlier models of
\cite{Barman01,Barman02} were constructed under the cond approximation and,
consequently, contained a larger number of free metals along with TiO and VO
molecular absorption compared to a rainout model.  These additional sources of
optical/UV opacity lead to a hotter upper atmosphere and a very shallow
photoionization depth for Na (only a 5\% reduction of Na across the region
probed by transit spectroscopy).  \cite{Fortney03} also explored Na
photoionization and found that their atmosphere model (which included rainout)
could be brought into reasonable agreement with the Na observations.  The
models presented here account for the angular dependence of ionziation on the
limb's dayside, but not on the night side. However, \cite{Fortney03} have shown
that Na ionization is still present at $\sim 5^\circ$ past the terminator for
pressures relevant to the transit spectrum modeled here.  

\begin{figure}[!t]
\plotone{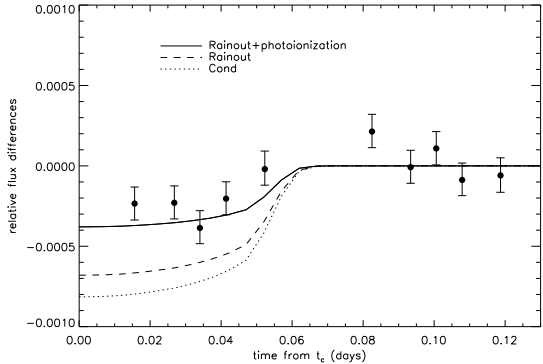}
\caption{
Relative flux differences (with time from center of eclipse) between a
wavelength band centered on the Na doublet and the mean of two wavelength bands
on either side of the Na doublet. The total wavelength range is indicated by
vertical lines in Fig. \ref{rlam1}. Each model is cloud-free with solar
abundances and $\alpha=0.25$. See Figure legend and text for the
distinguishing characteristics of each model. Symbols are the observations of
\cite{Charb02} with 1 $\sigma$ error bars.
\label{na1}}
\end{figure}

Figure \ref{rlam1} compares cloud-free solar abundance rainout models with and
without photoionization of Na and K.  While these models include
photoionization, the LTE approximation on the atomic level populations and line
source function is maintained.  Ionization (radially) at the limb reaches 50\%
at $P \sim 0.1$ and 1.2 mbar for Na and K, respectively, and stops at $P \sim$
2 and 7 mbar, respectively, resulting in a reduction of the \Rlam-peaks across
the Na and K lines. For K, this reduction extends out to the line-wings
resulting in a significantly smaller mean \Rlam, bringing the model into $\sim
3 \sigma$ agreement with the Knutson et al.  measurement.  The impact of Na
photoionization is mostly confined to the core of the doublet leading to only a
small reduction of the mean \Rlam, but sufficient to bring the model into $\sim
1 \sigma$ agreement with the observations at these wavelengths.  Though not
included here, ionization past the terminator onto the night side should
further improve the agreement across the Na and K doublets.  The two narrow
features on the red wing of the K doublet are due to Rb, which should also be
affected by photoionization due to its very low first ionization potential.
Photoionization of Rb resulted in a near complete removal of these features,
but reduced the mean \Rlam\ for this bin by less than 0.002 \Rjup.
 
The broad wavelength bins allowed Knutson et al. to obtain very precise
relative \Rlam\ measurements; however, as illustrated by the Na doublet, such
broad bins limits the constraints that can be placed on atomic absorption
features.  In contrast, the narrow range analyzed by \cite{Charb02} resulted in
much larger \Rlam\ error-bars, but still precise enough to easily distinguish
between the various models shown in Fig. \ref{rlam1}.  The flux differences
across the Na doublet as a function of time (measured from the transit center)
were computed for each model using the same narrow wavelength bins as
\cite{Charb02},  indicated in Fig.  \ref{rlam1} by vertical dashed lines. 
Fig. \ref{na1} compares model transit curves to the 2002 Na measurements and shows
a large discrepancy between the observations and solar abundance models with
pure equilibrium chemistry (cond or rainout).  Including photoionization brings
the baseline rainout model into rough agreement with the observations.  Note
the cond model shown in Fig. \ref{na1} does not include photoionization
(similar to the LTE model from \cite{Barman02}) and illustrates how far off the
predicted \Rlam\ can be across the Na doublet under simplified assumptions.

\begin{figure*}[!thb]
\plotone{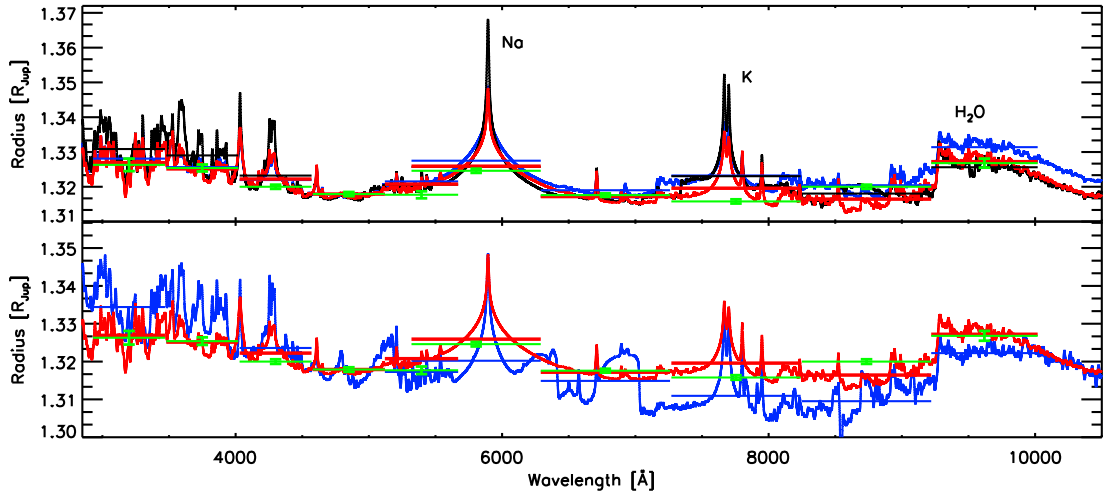}
\caption{
Monochromatic transit radii over the STIS spectral range.  Top panel: solar
abundance models, cond (black) and rainout(red), are compared to a model with
10$\times$ solar metal abundances including rainout (blue). Lower panel: solar
abundances baseline rainout model with $\alpha=0.25$ (red) compared to a
similar model with $\alpha=0.5$ (blue).  Horizontal bars have the same meaning
as in Fig. \ref{rlam1}.
\label{rlam2}}
\end{figure*}

\subsection{Rainout, Metallicity, and Temperature}
The rainout model used here reduces the individual metal abundances
with depth iteratively until clouds do not form, thus mimicking efficient
gravitational settling.  However, the removal of metals is not 100\% efficient,
leaving behind a variety of atoms (in addition to Na and K) to contribute to
the line opacity.  The impact of rainout on the planet's atmosphere is made
apparent by comparing rainout and cond models (Fig. \ref{rlam2}). In
the cond model grain opacities are simply ignored in the transfer equation
while the chemistry remains that of a pure equilibrium model without any actual
removal of refractory elements.  Thus, the cond model represents the minimum
impact of grain formation on the stratified abundances and leaves behind a
considerable amount of free metals along with molecules like TiO and VO.  This
leads to much stronger atomic lines, including Na and K along with TiO and VO
bands.

Apart from a very minor contribution from molecules, the \Rlam\ features at
$\lambda < 0.8 \mu$m are all due to atomic line opacity
\footnote{\cite{Ballester07} claim the \Rlam\ rise in the blue/UV is due to the Blamer
edge of hot hydrogen in an extended atmosphere. While the present model does
not support this explanation, if verified, it could indicate that rainout is
more efficient than predicted here.}. These narrow features are due to blended
lines from metals like Ca, Al, Fe, Ni, Mn, and Cr. Note that, while Rayleigh
scattering does contribute to the opacity in the blue/UV, a removal of atomic
line opacity would drop \Rlam\ by $\sim 0.02$\Rjup\ for $\lambda < 0.45 \mu$m.
The top panel of Fig.  \ref{rlam2} compares the solar abundance rainout model
with photoionization (shown in Fig.  \ref{rlam1}) to a rainout model with
10$\times$ solar abundances.  Since the metal abundances were uniformly scaled,
the grain formation and corresponding removal of refractory elements were also
uniformly enhanced.  The additional opacity altered the T-P profile as well as
the total atmospheric extension.  These factors contribute to a similar \Rlam\
pattern in the optical but enhanced molecular absorption features in the
red/near-IR.  The increase in metals increases the \h2o\ absorption feature
along with the wings of the Na and K doublets (which form deeper in the
atmosphere where photoionization is less important).  Also, FeH and CrH
absorption is larger in the metal-rich atmosphere leading to an increase in
\Rlam\ on both sides of the K doublet.  The metal-enhanced \Rlam\ spectrum is
in good agreement with the observations near 0.9$\mu$m, but is noticeably too
high across both the K doublet and the \h2o\ band.

Transit spectroscopy can also help constrain the temperatures across the limb.
The lower panel of Fig. \ref{rlam2} compares an $\alpha=0.5$ model (i.e.  a
model with a hotter dayside-like T-P profile) to the cooler $\alpha=0.25$
baseline model.  In the hotter model, grain formation is less pronounced
resulting in a greater concentration of most free metals and, thus, stronger
UV/blue absorption lines than in the $\alpha=0.25$ model.  Higher temperatures
at depth also result in equilibrium concentrations of Na and K that are several
times lower than found at the cooler temperatures of the $\alpha=0.25$ model.
The net results are two distinctively different \Rlam\ spectra with the
$\alpha=0.25$ model being more consistent with the Knutson et al.
measurements.

\section{Summary}
Photoionziation plays an important role for both Na and K, and potentially many
other species.  No evidence is found to support a large metal enhancement
(e.g., 10$\times$ solar), though smaller Jupiter-like enhancements are not
ruled out.  Furthermore, the agreement between model and observations
demonstrated here alleviates the need for substantial cloud coverage along the
limb between ~ 0.05 and 0.001 bars, which would otherwise truncate and flatten
the \Rlam\ spectrum \cite[]{Brown01}.  In the baseline model, the predicted
location of clouds (e.g., MgSiO$_3$) lies just below the minimum \Rlam ($\sim
1.315$\Rjup) across the STIS wavelength range, consequently deep clouds (and
also very high clouds) remain a possibility. 
 
The models presented here also predict \Rlam\ variations ($\sim 0.02$ \Rjup)
across the 2-0 R-branch of CO in the $K$-band.  This is inconsistent (at
2.5$\sigma$) with a non-detection of CO based on Keck-NIRSPEC transit spectra
taken in 2002 \cite[]{Deming05b}.  It is likely that using a single T-P profile
to represent the horizontal and pole-to-pole variations across the limb
averages out too much of the upper atmospheric structure; this simplification
might explain the $K$-band discrepancy. High clouds may also be involved.  In
addition, the Keck and HST observations were taken between 1 and 2 years apart
and time-dependent atmospheric variations along the limb cannot be ruled out
\cite[]{Menou03}.

While \cite{Knutson07} did not attribute their measured \Rlam\ variations to
any absorption features (this was not the focus of their paper), the models
presented above clearly show that these measurements are consistent with strong
water absorption near 1 $\mu$m.   A detection of water in the limb of HD209458b
is nominally at odds with a recent {\em Spitzer} IRS spectrum that shows no
\h2o\ features for this planet\cite[]{Richardson07}.  These data were taken
during {\em secondary} eclipse and directly probe the planet's dayside with
negligible contribution from the limb.  It is possible that the dayside
atmosphere is nearly isothermal \cite[]{Fortney06} which would result in a
spectrum with no detectable water absorption features, despite a copious water
supply.  The transmission spectrum, however, would contain absorption features
independent of an isothermal dayside or limb.  

\acknowledgements
The author thanks B. Hansen and the referees for their comments.  This research
was supported by NASA Origins and Spitzer Theory Grants and made use of NASA's
Project Columbia computer system.

\end{document}